\def\deg{\hbox{$^\circ$}}                           
 
\def\gray{\mbox{$\gamma$-ray}}                      
\def\grays{\mbox{$\gamma$-rays}}                    
\def\apj{ApJ}                                       
\def   \ni {\noindent}

\def   \ssk {\vskip  5truept}

\def   \bsk {\vskip 15truept}
 
\def   \newpage {\vfill\eject}
\def   \newline {\hfil\break}

\documentstyle[epsfig]{article}
\begin{document}

\hsize 5truein
\vsize 8truein
\font\abstract=cmr8
\font\keywords=cmr8
\font\caption=cmr8
\font\references=cmr8
\font\text=cmr10
\font\affiliation=cmssi10
\font\author=cmss10
\font\mc=cmss8
\font\title=cmssbx10 scaled\magstep2
\font\alcit=cmti7 scaled\magstephalf
\font\alcin=cmr6 
\font\ita=cmti8
\font\mma=cmr8
\def\ref{\par\noindent\hangindent 15pt}
\null


\title{\ni Possibility of the detection of classical novae with the shield of the INTEGRAL-spectrometer SPI}                                               

\bsk \bsk
\author{\ni P.~Jean$^{1,2}$, J.~G\'omez-Gomar$^2$, M.~Hernanz$^2$, J.~Jos\'e$^2$,  J.~Isern$^2$, G.~Vedrenne$^1$, P.~Mandrou$^1$, V.~Sch\"onfelder$^3$, G.G.~Lichti$^3$, R.~Georgii$^3$}                                                       
\bsk
\affiliation{$^1$Centre d'Etude Spatiale des Rayonnements, CNRS/UPS, 9 avenue du colonel Roche, 31028 Toulouse, France; $^2$Institut d'Estudis Espacials de Catalunya (IEEC), Edifici Nexus-201, C/ Gran Capit\`a 2-4, E-08034 Barcelona, Spain; $^3$Max-Planck-Institut f\"ur extraterrestrische Physik, Postfach 1603, 85740 Garching, Germany 
}                                                
\bsk
\baselineskip = 12pt

\abstract{ABSTRACT \ni The shield of the INTEGRAL spectrometer provides a large detection area with a wide field-of-view. Calculations have been performed to check whether the temporal analysis of the counting rate of the SPI anticoincidence allows the detection of explosions of novae. The background rate of the shield as well as its response to gamma-ray have been modelled with monte-carlo simulations. Accounting for uncertainties in the rate of novae, their distribution in the Galaxy and their light curves in hard X-ray domain, the number of nova explosions detectable with this method during the INTEGRAL mission, is estimated. Such observationnal mode will allow to improve our knowledge on nuclear runaway in novae. Since the maximum of magnitude in the visible happens later than in gamma-ray, SPI will provide alert for optical observations.}                                                    
\bsk
\baselineskip = 12pt

\bsk
\baselineskip = 12pt


\text{\ni 1. INTRODUCTION
\ssk
\ni 
Observation of \grays\ emitted by novae is a challenge for gamma-ray astronomers. Such a detection would provide new insights in understanding of explosive nucleosynthesis as well as galactical abundance of elements. Presently, only upper-limits of \gray\ emission of novae have been provided by space-borne instruments. The high resolution germanium spectrometer (SPI) of the future INTEGRAL observatory is designed for the detection of astrophysical \gray\ lines. The detection plane is made of 19 Ge detectors and is surrounded by an active shield made with 90 bismuth germanate (BGO) scintillators, in order to reduce the Ge instrumental background. The BGO-shield provides a large detection area and a wide field-of-view, that could be used not only to reduce instrumental background of the spectrometer but also to detect \gray\ sources. We present here a study of the capability of the SPI BGO-shield for the detection of classical novae by analysing temporal fluctuations of its counting rate. 

\bsk
\ni 2. GAMMA-RAY EMISSION OF NOVAE 
\ssk
\ni 
During a nova explosion radioactive nuclei are synthesized in the envelope. Their decay leads to the emission of \grays\ that can escape the ejecta depending on
\newpage
\ni
their opacity condition. G\'omez-Gomar et al. (1998) calculated \gray\ spectra and light curves of different types of novae with a complete hydrodynamical code for the estimation of the velocity, the temperature profile and the yield of radioactive nuclei, and used a Monte-Carlo code to treat the transport of $\gamma$-ray through the envelope. Figure 1 shows the simulated light curve for 3 types of novae.
\begin{figure}
\centerline{\psfig{file=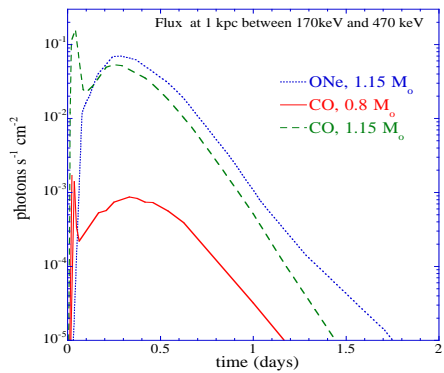,height=5.0cm,width=8.cm}}
\caption{FIGURE 1. Evolution of the \gray\ emission for different models.}
\end{figure}

\bsk
\ni 3. SPI SHIELD RESPONSE TO A GAMMA-RAY FLUX
\ssk
\ni 
A fraction of the \gray\ flux emitted by a nova will interact with the BGO-shield of SPI. It will induce a counting rate that depends on the \gray\ spectrum and the direction of the nova with respect to the shield. The effective detection area of the shield has been estimated as a function of azimuth- (A) and zenith-angles (Z), and the energy of $\gamma$-ray, with Monte-Carlo simulations using the GEANT code and an accurate model of the INTEGRAL observatory (Sturner, 1998). The orientation is such that the z-axis is the spectrometer pointing-axis and the x-axis direction is opposite to the IBIS instruments. The effective area as a function of Z and the \gray\ energy is presented figure 2 for A=60\deg.  

\begin{figure}
\centerline{\psfig{file=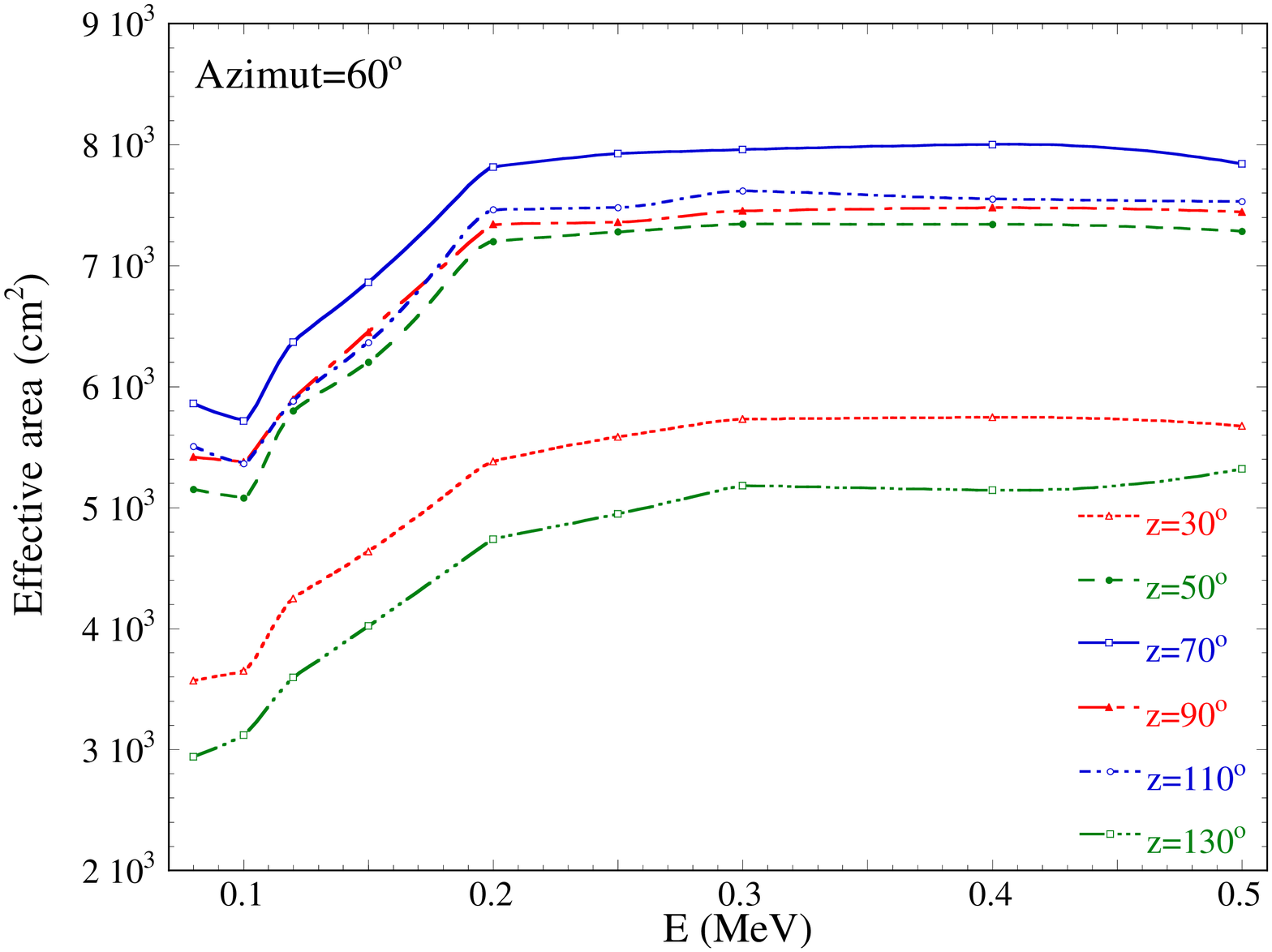,height=4.0cm,width=7.cm}}
\caption{FIGURE 2. SPI shield effective area at A=60\deg\, versus energy for several zenith-angle values.
}
\end{figure}


\bsk
\ni 4. SHIELD BACKGROUND RATE 
\ssk
\ni 
Under space conditions, the satellite and the instrument aboard will be irradiated by high-energy particles leading to a background rate in the BGO-shield. The calculation of the rate induced by cosmic-ray particles (p$^+$, $\alpha$, e$^-$), cosmic diffuse $\gamma$ and internal radioactive decays, has been performed with Monte-Carlo simulations using the GEANT code. Figure 3 shows the results. The total rate obtained is $\approx$5.5 10$^4$ counts s$^{-1}$ in solar maximum period.
\begin{figure}
\centerline{\psfig{file=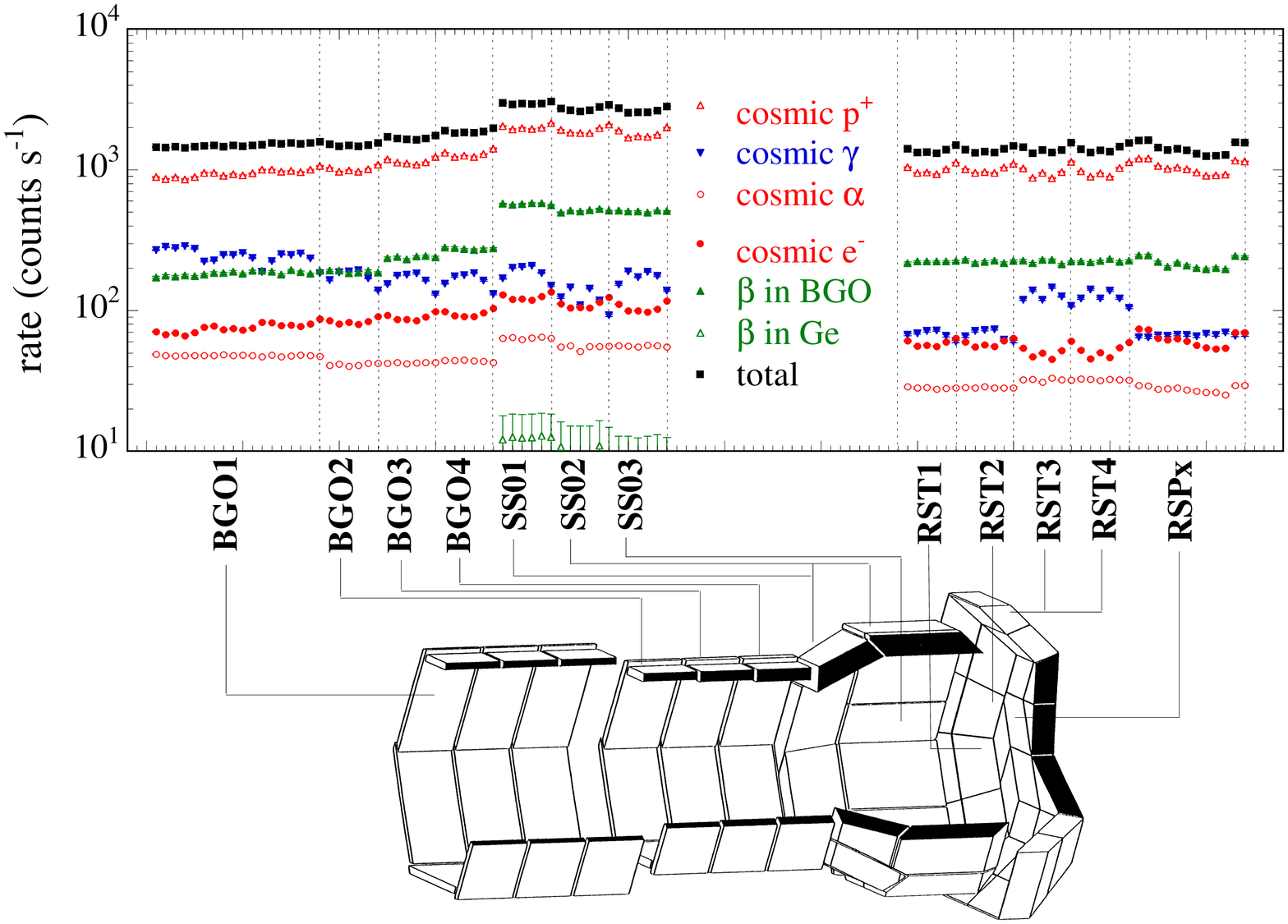,height=5.0cm,width=8.cm}}
\caption{FIGURE 3. Estimation of the rates induced by cosmic-ray particles in the BGO-blocks.
}
\end{figure}
\bsk
\ni 5. SIMULATION OF THE OBSERVATION 
\ssk
\ni 
Using previous calculations, the SPI-anticoincidence rate as a function of time has been estimated for novae at different distances and directions. This rate was compared with the normal background counts for different time binnings and the statistical significances of the signal has been calculated for a day. When an enhanced number of counts for the time bin in question exceeds the 5$\sigma$ level, we performed a statistical test to determine the probability that the 'observed' temporal-serie of rate is due to background fluctuations. When this probability was 5$\sigma$, it was assumed that this rate increase was induced by a nova explosion. The maximum distances for such detections have been calculated as a function of the direction for 3 types of novae. An example is presented in figure 4 for the ONe nova model.
\begin{figure}
\centerline{\psfig{file=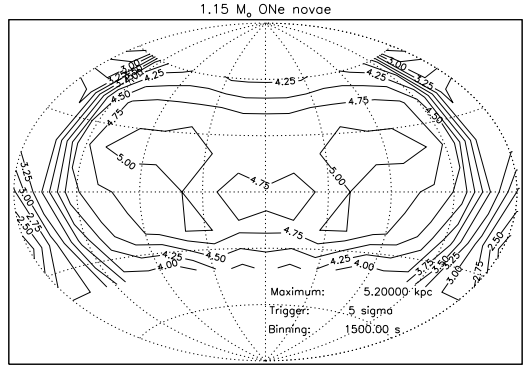,width=7.cm}}
\caption{FIGURE 4.  Maximum distances (kpc) for the detection of a 1.15 M$_{\odot}$ nova as a function of its direction.}
\end{figure}
\bsk
\ni 6. CONCLUSION 
\ssk
\ni 
Using 2-hour binning, the maximum distances for detections at 5$\sigma$ level of novae are 7.4 kpc, 0.7 kpc and 8 kpc for 1.15 M$_{\odot}$ ONe, 0.8 M$_{\odot}$ CO and 1.15 M$_{\odot}$ CO type novae respectively. Assuming that the infrared emission of our Galaxy is a tracer of the old star population and therefore of novae, the frequency of novae closer than a given distance can be estimated using the distribution of Kent, Dame \& Fazio (1991). Since $\approx$30\% of novae might be ONe novae, we can expect to detect $\approx$3 events per year with the proposed observation mode if we assume that their $\gamma$-ray emission is similar to the 1.15 M$_{\odot}$ model.
The analysis of the SPI-shield rate would inform us of a nova explosion somewhere in the Galaxy. A localization of such an event would be possible by comparing the counting rate in individual BGO-blocks. The counting rate increase of scintillators oriented to the direction of the nova would be larger than the others. 
Although INTEGRAL will spend more than 90\% of time per orbit out of the outer electron radiation belt, we can expect background rate variations. These would reduce the sensitivity of such a mode of detection. However, we can minimize this effect with a model of the shield-rate that uses the radiation monitor data.
The intensity and temporal profile provided with such observational mode will allow our knowledge of nuclear runaway in novae to be improved. Moreover, since the maximum of magnitude in the visible happens later than in $\gamma$-ray, SPI would be able to provide alerts for optical observations. The proposed mode of detection can also be used for detection of others $\gamma$-ray transients that have significative emission above 0.1 MeV.
\bsk
}

\ssk
\baselineskip = 12pt
{\abstract \ni ACKNOWLEDGMENTS
Research partially supported by the training and Mobility Researchers Programme. Access to large installation, under contract ERBFMGECT050062 - Access to supercomputing facilities for european researchers established between the European Community and Community and CESCA-CEPBA. The spectrometer SPI is supported by the german government through DLR grant 50OG9503 0.

}

\bsk
\baselineskip = 12pt

{\references \ni REFERENCES
\ssk
\ref Gomez-Gomar, J., Hernanz, M., Jose, J., Isern, J., 1998, MNRAS, 295, 1-9
\ref {Kent}, S.~M., {Dame}, T.~M., \& {Fazio}, G. 1991, \apj, 378, 131-138
\ref Staubert, R., 1985, 19th Int. Cosmic Ray Conf. Papers, {\it 3}, 47 
\ref Sturner, S., 1998, private communication
}                      

\end{document}